\documentclass[aps,prl,twocolumn,showpacs,amsmath,a4paper,amssymb,10pt]{revtex4-1}
\usepackage{graphicx}
\usepackage{CJK}
\usepackage{color}

\begin{document}
\begin{CJK*}{UTF8}{gbsn}

\title{Two-Dimensional Attosecond Electron Wave Packet Interferometry}

\author{Xinhua Xie (谢新华)}
\email[Electronic address: ]{xinhua.xie@tuwien.ac.at}
\affiliation{Photonics Institute, Vienna University of Technology, A-1040 Vienna, Austria}

\date{\today}

\begin{abstract}
We propose a two-dimensional interferometry based on electron wave packet interference with a cycle-shaped orthogonally polarized two-color laser field.
With such method, sub-cycle and inter-cycle interferences can be disentangled into different direction in the measured photoelectron momentum spectra.
With the cycle-shaped laser field, the Coulomb influence can be minimized and the overlapping of interference fringes with the complicate low-energy structures can be avoided as well.
The contributions of excitation effect and long-range Coulomb potential can be traced in the Fourier domain of the photoelectron distributions.
With these advantages, it allows to get precise information on valence electron dynamics of atoms or molecules with attosecond  resolution and additional spatial information with angstrom resolution.
\end{abstract}

\maketitle
\end{CJK*}


The dynamics of valence electrons is the key to understand most ultrafast processes, from ionization and excitation of atoms and molecules, to the dissociation and isomerization reaction of molecules\cite{krausz09}.
Therefore exploring valence electron dynamics in atoms and molecules is one of the most essential tasks of ultrafast science.
The natural timescale of valence electron motion is on the subfemtosecond (or attosecond) timescale.
Thus to get insight into the dynamics of valence electrons, techniques with attosecond temporal resolution are demanded.
Several methods, such as attosecond extreme ultraviolet or X-ray spectroscopy\cite{Gulielmakis2010,he10,chini12,loh13}, high harmonics spectroscopy\cite{olga09:co2,shiner11,worner11}, and photoelectron spectroscopy based on electron wave packet (EWP) interferences\cite{remetter06,gopal09,huismans11,xie12,Boguslavskiy2012}, already showed the ability and their advantages on study attosecond electronic dynamics.
Not only attosecond temporal resolution but the accessibility to the bound electronic phase and structure is required for probing the motion of valence electrons.
Since the phase and structure information of the bound electronic states is directly encoded in the released EWPs, such information can be retrieved from the interference fringes of EWPs in photoelectron spectra.
The biggest challenge of photoelectron spectroscopy is to disentangle different contributions from the interference fringes and read out corresponding information, because the fringes induced by different kinds of interferences are in general mixed with one another in the photoelectron spectra.

\begin{figure}[htbf]
\centering
\includegraphics[height=0.24\textwidth,angle=0]{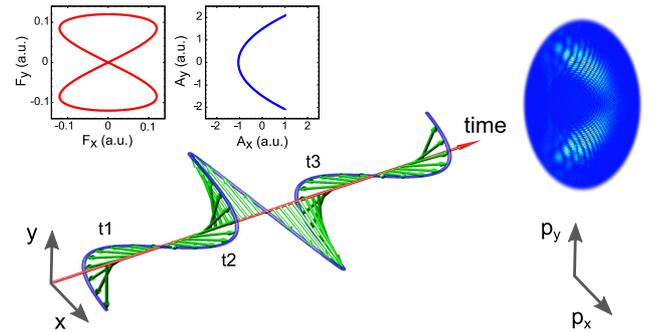}
\caption{Schematic view of the 2D-EWPI.
The evolution of the cycle-shaped OTC electric field is presented as the blue curve over time, while the green arrows indicate electric field vectors at corresponding time.
EWPs released at different time (t1,t2,t3) will end at the same final momenta and interfere with one another in the momentum space.
The 2D spectra on the right side is a momentum distribution solution of the time-dependent Schr\"odinger equation.
The insets on the left upper corner present the electric field and the vector potential of the cycle-shaped OTC laser field in the laser polarization plane. } \label{fig:peda}
\end{figure}

The EWP interferometry is based on the interference of EWPs, which are released through tunneling ionization when a strong laser field acts on an atom or a molecule (Fig.~\ref{fig:peda}).
In general, the released EWPs carry electron or nuclear dynamics and the orbital structure information of the system.
Therefore, the EWPs interference can be used to retrieve the information of the molecular orbital\cite{meckel08}, ionization dynamics \cite{xie12} and the influence of the ionic Coulomb potential\cite{huismans11}.

In brief, there are three most general types of EWP interferences in strong field regime.
The first type is inter-cycle interference (ICI) with EWPs released from different laser optical cycles which leads to ATI(above-threshold ionization)-like structure in momentum distribution\cite{arbo10,Boguslavskiy2012}.
The second type is sub-cycle interference (SCI) with EWPs from different half cycle within one optical cycle\cite{linder05,xie12,arbo06,arbo14}.
The third type interference is produced by EWPs from the same quarter of the field with scattering on the ionic potential\cite{huismans11,spanner04}.
A precisely defined momentum-to-time mapping and the correct assignment of interference fringes caused by different kind interferences is necessary for retrieving the information of electronic dynamics from the measured interference pattern.
Previous experiments show that the treatment of Coulomb influence is important but not trivial\cite{huismans11,xie12,xie13,zhang14_2}.
Besides, because of the mixture of different kind interferences, the fringe positions of one kind interference will be modified by the other kinds which will thus affect the precision of the phase reconstruction\cite{arbo10}.
Another important effect is the influence from the complex low-energy structures (LES)\cite{blaga09:les,quan09,liu09,Dimitrovski14}.

To achieve high temporal resolution, a well-identified interference pattern is required for the study of the ultrafast electron dynamics.
In this letter, we propose a two-dimensional EWP interferometry (2D-EWPI) with a cycle-shaped orthogonally polarized two-color (OTC) laser field.
It leads to a well-defined 2D interference pattern in the momentum distribution of photoelectrons.
Previously, OTC fields were proposed and applied to control electron recollision\cite{kitzler05,kitzler07}, image wave functions based on the generation of high harmonics\cite{kitzler08,shafir09} and control electron emission and correlation in the atomic single and double ionization\cite{zhang14,zhou11,zhang14_2}.
With the 2D-EWPI using OTC fields, the difficulties mentioned previously can be easily overcome.
The fringes induced by different kinds of interferences appear along different directions in the 2D momentum spectra.
Different type of structures can be straightforward filtered out in the Fourier transformed frequency domain.
To compare with linearly polarized laser field driven interferometry, the biggest advantage of the 2D-EWPI is the disentanglement of ICI and SCI which are mixed with each other in case of using linearly polarized fields.
With the OTC field, the influence of Coulomb potential is minimized as well because the main part of the electron wave packet will miss the ionic potential core.
Additionally, we can avoid overlapping with complex LES because the sub-cycle interference structure locates away from zero momentum.
Yet another important advantage is the structure caused by electronic excitation can be distinguished as well.

To avoid overlapping between the interference fringes and the complex LES, the final momentum distribution should have non-zero value.
To achieve such goal, the most straightforward way is to use elliptically polarized driven laser pulses.
With elliptical polarized laser pulses, the Coulomb effect can be minimized because the main part of released EWPs will not scatter on the ionic potential core.
However, with elliptical field, the SCI will disappear because EWPs released within one optical cycle will not interfere in the momentum space.
To overcome all difficulties mentioned previously, in this work, we choose a cycle-shaped OTC laser field from superposition of a fundamental field and its phase-locked orthogonally polarized second harmonic.
We fix the phase difference between the two fields to 0.5$\pi$.
The electric field and the vector potential of such superimposed field is sketched in Fig.~\ref{fig:peda}.
The vector potential has a bow-like shape in the polarization plane.
 It avoids overlapping with the LES region, and within one fundamental optical cycle the vector potentials of the different half optical cycles overlap exactly with each other which leads to SCI of the EWPs in the momentum space.

\begin{figure}[htbf]
\centering
\includegraphics[width=0.46\textwidth,angle=0]{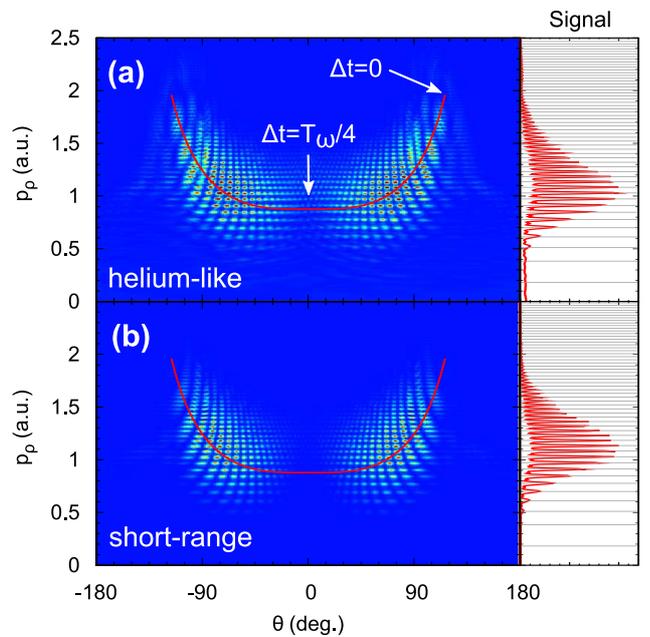}
\caption{Simulated 2D EWP interference pattern for helium-like model (a) and short-range potential model (b) in the laser polarization plane with the polar coordinate. Red line indicates the vector potential of the cycle-shaped OTC field. The small panels on the right side present the integrated spectra over $p_\rho$ with horizontal gray lines indicating the positions of ATI-like peaks.} \label{fig:2d-mom}
\end{figure}

By numerical solving the 2D time-dependent Schr\"odinger equation with the single active electron approximation\cite{xie07jmo,xie07pra}, we simulate photoelectron spectra from single ionization of an atom in a cycle-shaped OTC field.
We employ a helium-like atomic model with a soft-core potential $V(x,y)=-1/\sqrt{x^2+y^2+a^2}$, where the screening parameter $a=0.308$ to reproduce the helium ionization potential 0.90 a.u..
The electric field of the OTC field is defined as $\vec{F}(t)=f(t)(\cos(2\omega + 0.5\pi){\bf \hat{\bf x}}+\cos(\omega t){\bf \hat{\bf y}})$, where $f(t)$ is a super-Gaussian envelope function to ensure a flat-top laser pulse, and $\omega=0.057$ a.u. (corresponding to 800 nm in wavelength) is the center frequency of the fundamental field.
A 3D view of the OTC field in the polarization plane over time is displayed in Fig.~\ref{fig:peda}.
We use laser peak intensity 3.5$\times 10^{14}$ W/cm$^2$ and pulse duration 9.2 fs (full width at half maximum) for both the fundamental and the second harmonic pulses in the simulations.
With the super-Gaussian envelope, there are three optical cycles of the fundamental field within the pulse.

To figure out the influence of the long-range Coulomb potential and the excitation effect on the EWP interference, we perform simulations with a short-range potential model.
The short-range potential is defined as $V(x,y)=-e^{-2\ln2(x^{12}+y^{12})/R_0^{12}}/\sqrt{x^2+y^2+a^2}$.
By varying the radius parameter $R_0$, we can truncate the long-range Coulomb potential tail and control the number of bound states and the energy gap between bound states of the model.

Fig.~\ref{fig:2d-mom} (a) presents the simulated 2D momentum distribution of photoelectron in the polar coordinate of the laser polarization plane for the helium-like model.
There are two kinds of patterns from the two types of interferences which can be easily distinguished in the 2D spectra.
ATI-like peaks induced by ICI are shown along the vertical direction with no dependence on angle and the structure caused by the SCI in the horizontal direction follows the red line given by the vector potential of the laser field.
With the relation between the vector potential at EWP birth time $t_b$ and the electron final momentum $\vec{p}_e=-\vec{A}(t_b)$, the time-to-momentum mapping can be determined, as the time difference shown in the figure.
Because of the Coulomb influence is insignificant and no overlapping with the LES, a precisely defined mapping between final momentum and ionization time can be obtained.
With such mapping, electron dynamics during field ionization can be retrieved, as discussed in Ref.~\cite{xie12}.
Meanwhile, the involved electronic state can be determined from sub-cycle interference, since the positions of ATI-like peaks are determined by the laser frequency, the ionization potential ($I_p$) of the electronic state and the ponderomotive energy ($U_p$) of the field, with a relation $E_{ATI}=n\omega-I_p-U_p$.
In Fig.~\ref{fig:2d-mom}, the momentum distribution along $\rho$ direction shows the ATI-like peaks which perfectly agree with the prediction by the relation. It means such structure can be employed to determined the ionization potential, i.e. the energy, of the bound electronic state\cite{Boguslavskiy2012}.
One the other hand, with a well-known target, we know the ionization potential such that the laser intensity can be calibrated from the relation.

The simulated 2D momentum spectra for the short-range potential is shown in Fig.~\ref{fig:2d-mom} (b).
The parameter $R_0=2$, with which there is only one deeply bonded state with energy -0.90 a.u..
Therefore, the excitation effect will be strongly suppressed.
To compare with Fig.~\ref{fig:2d-mom} (a), we notice that the overall structures are similar, except that there are more structure near the $\Delta t=0$ region for the simulated spectra of helium-like model.
Such structures originate from excitation effect when the electric field is very weak\cite{xie12}.

\begin{figure}[htbf]
\centering
\includegraphics[width=0.46\textwidth,angle=0]{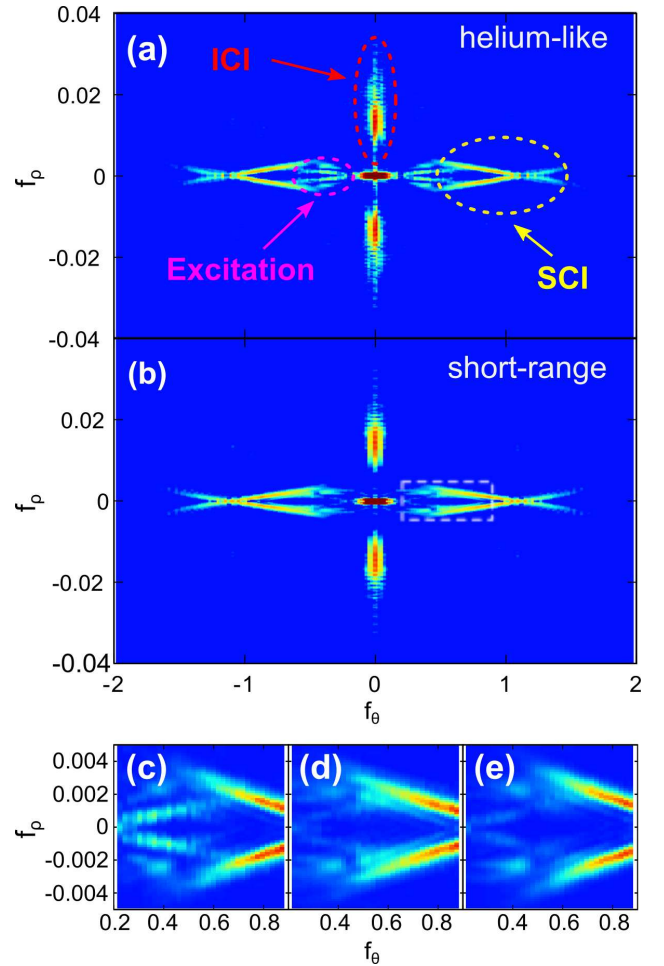}
\caption{Fourier transform of the 2D photoelectron spectra in Fig.\ref{fig:2d-mom} for the helium-like model (a) and the short-range potential model (b). As marked in panel (a), different frequency regions originate from different kinds of EWP interferences. Panels (c-d) present a zooming of the 2D photoelectron spectra in the rectangle region as shown in the panel (b) for the helium-like model (c), the short-range model (d) and the helium-like model with projection off excited states (e).} \label{fig:2d-mom-fft}
\end{figure}

To get insight to different kinds of fringes and the difference between the helium-like model and the short-range potential model, we perform 2D Fourier transform of the simulated photoelectron spectra.
As shown in Fig.~\ref{fig:2d-mom-fft}, different kind interference structures and contributions to the interference fringes present as isolated distributions in the Fourier transformed photoelectron spectra.
The ICI structure appears along $f_\rho$-coordinate at $f_{\theta}$=0 within the red ellipse in Fig.~\ref{fig:2d-mom-fft}(a), while SCI presents as scissors-like structure along $f_{\theta}$-coordinate within the yellow ellipse in Fig.~\ref{fig:2d-mom-fft}(a).
By comparing the Fourier transformed spectra between the helium-like model and the short-range potential model, we locate the structure induced by the excitation effect also along $f_{\theta}$-coordinate but with small $|f_{\rho}|$ within the magenta ellipse in Fig.~\ref{fig:2d-mom-fft}(a).

To further confirm the structure induced by the excitation effect, we compare the structures between the helium-like model and two models with suppressed excitation effect.
One model is the short-range potential model which is already shown. The other model is implemented the same as the helium-like model but with projecting away the excited 2\textit{p} and 3\textit{p} states at each step of time propagation to get rid of the contribution of excitation from those states.
The results of the Fourier transformed photoelectron spectra near the excitation structure are shown in Fig.~\ref{fig:2d-mom-fft}(c), (d) and (e) for the helium-like model, the short-range model and the helium-like model with projection off excited states, respectively.
It is obvious that the signals of excited induced structure in the left side of the figures are suppressed for the short-range model and the helium-like model with projection of excited states.

In the simulation, we carry out a set of simulations with increasing cutoff position of the short-range potential by varying the radius parameter $R_0$.
We find that the long-range Coulomb potential induces shifting of the SCI fringes along the vector potential direction.
Also, we note the changes on the excitation induced fringes because more excitation states are included with increasing the long-range Coulomb potential tail.
Meanwhile, the ICI fringes have not significant changes.

\begin{figure}[htbf]
\centering
\includegraphics[width=0.48\textwidth,angle=0]{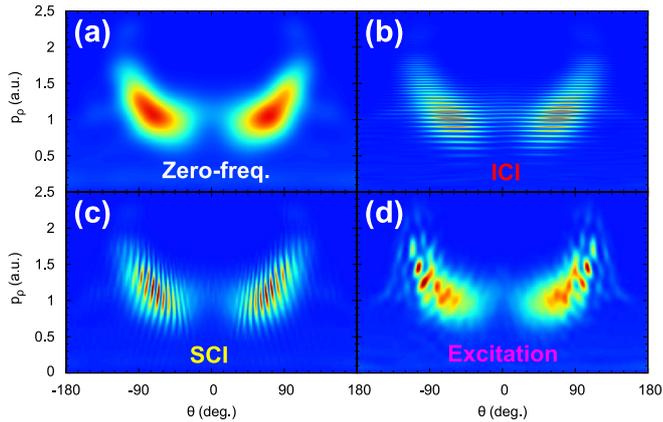}
\caption{Fourier frequency filtered 2D photoelectron spectra of the he-like model: (a) spectra with zero-frequency, (b) ICI induced spectra, (c) SCI induced spectra, (d) Excitation induced spectra. } \label{fig:2d-mom-fft-filtered}
\end{figure}

Since the contributions from different kinds of interferences can be separated in the frequency domain, we filter out different region in frequency domain and Fourier transform back the spectra to the momentum space.
The results are presented in Fig.~\ref{fig:2d-mom-fft-filtered} for the helium-like model and the short-range potential model.
The zero-frequency spectrum is the overall shape of a momentum distribution which contains the amplitude information of the interfered EWPs.
The ICI spectrum includes only the interference of EWPs with time interval equaling the integer multiples of the optical cycle.
The SCI spectrum is induced by EWPs released within one optical cycle.
The excitation spectrum tells the information of excitation effect before the EWPs releasing.

Because of the disentanglement of different contributions to the 2D momentum distribution, we can use a formula to describe the interference pattern, $S(p_\rho,\theta)=I_0(p_\rho,\theta)\cos^2\frac{\Delta\Phi_{I}}{2}\cos^2\frac{\Delta\Phi_{S}}{2}\cos^2\frac{\Delta\Phi_{E}}{2}$, where $\Delta\Phi_{I}$, $\Delta\Phi_{S}$ and $\Delta\Phi_{E}$ are the relative phases corresponding to ICI, SCI and excitation effect, respectively.
From the measured interference pattern, the amplitude and the phase of the three types of interference can be retrieved from the positions, the amplitudes and the visibility of the fringes\cite{xie12}.

To get the dynamics of a quantum system, the phase information of the system is most important.
To retrieve the phase of a quantum system from the measured interference pattern.
The relation between the measured quantity and the intrinsic phase of the system should be determined.
Since the interfered EWPs undergoes ionization process and acceleration process in the combined field of laser electric field and Coulomb field before the formation of interference pattern, the phase of the EWPs can be separated in to three terms as $\Phi_{S}=\Phi_b+\Phi_f+\Phi_c$, where $\Phi_b$, $\Phi_f$ and $\Phi_c$ represent the contribution from the bound state, electric field and Coulomb field, respectively\cite{xie12}.
The phase from the bond state $\Phi_b$ is determined by the energy of the bound system.
Since the laser field shape is well defined, the phase induced by the field and the Coulomb field can be calculated.
With the time-to-momentum mapping, attosecond dynamics of the system can be retrieved.

The Spatial information of the wave packet is encoded in the interference pattern as well.
First of all, OTC field vector is time-dependent, which means that electron is detached from different direction at different ionization time.
Moreover, the initial lateral momentum distribution will also be mapped to the final momentum distribution.
For an oriented molecule, the spatial phase information at the time of ionization can be well transferred to the interference pattern through lateral momentum.
For SCI and excitation spectra, there are two dimensions in the distribution.
One dimension is along the vector potential, which represents time difference between the two EWPs.
The other dimension is perpendicular to the vector potential which contains the information of initial lateral momentum during tunneling ionization.
It can in the end transformed into spatial space from the momentum space if the phase and amplitude can be retrieved from the interference pattern.
Because the momentum distribution is on the atomic unite scale, the spatial information determined by Fourier transformation has angstrom spatial resolution\cite{huismans11}.

Before closing, we shortly summarize the capability of the 2D-EWPI.
The ATI-like spectra can be used to investigate the ionization potential of the system or ionization contributions from different electronic states or molecular orbitals\cite{Boguslavskiy2012}.
It can be used to retrieved long time (longer than the optical cycle) scale dynamics.
The SCI spectra can be used to study the sub-cycle electron dynamics with attosecond resolution\cite{xie12}.
The excitation spectra can be specifically used to explore the time-dependent excitation effect during the strong field interacting with atoms and molecules.

In conclusion, we propose a 2D-EWPI with a cycle-shaped OTC laser field.
With such method, the different types of interferences can be well disentangled into different directions in the measured 2D photoelectron momentum spectra, which can be effectively analyzed with Fourier filtering analysis.
With the cycle-shaped laser field, the Coulomb influence can be minimized and the overlapping of interference structures with complicate LES can be avoided.
With these advantages, it allows to get precise temporal and spatial information from the measured interference fringes, which can be used to study electronic dynamics and structure of atoms or molecules with attosecond temporal resolution and angstrom spatial resolution.
The 2D-EWPI and the method used to analysis the spectra can be in general applied to more complex quantum systems.
Due to that the vector potential scales with the wavelength with fixed field strength, the 2D-EWPI favors longer wavelength driven laser field.
Additionally, it is compatible with pump-probe technique and gain more information from scanning target orientation and laser parameters as well.

This work was financed by the Austrian Science Fund (FWF), Grants No. P25615-N27. Thanks to Vienna Scientific Cluster (VSC) for providing computing resource under project Nr.70458.

\end{document}